\documentclass{iau} 
\usepackage{graphicx}
\usepackage{natbib}
\usepackage{amssymb}
\usepackage{amsmath}
\usepackage[bookmarks=true,         
     pdfnewwindow=true,      
     colorlinks=true,    
     linkcolor=cyan,     
     citecolor=cyan,     
     filecolor=cyan,  
     urlcolor=cyan,      
 ]{hyperref}

\title[Bayesian discrimination of the SED modelings of galaxies] 
{Bayesian discrimination of the panchromatic spectral energy distribution modelings of galaxies}

\author[Yunkun Han]   
{
	Yunkun Han$^{1,2,3}$
	Zhanwen Han$^{1,2,3}$
	Lulu Fan$^{4}$
}

\affiliation{
$^1$Yunnan Observatories, Chinese Academy of Sciences, 396 Yangfangwang, Guandu District, Kunming, 650216, P. R. China \\email: {\tt hanyk@ynao.ac.cn}\\[\affilskip]
$^2$Center for Astronomical Mega-Science, Chinese Academy of Sciences, 20A Datun Road, Chaoyang District, Beijing, 100012, P. R. China \\[\affilskip]
$^3$Key Laboratory for the Structure and Evolution of Celestial Objects, Chinese Academy of Sciences, 396 Yangfangwang, Guandu District, Kunming, 650216, P. R. China\\[\affilskip]
$^4$Institute of Space Science, Shandong University, Weihai, 264209, China
}

\pubyear{2018}
\volume{341}  
\jname{PanModel2018: Challenges in Panchromatic Galaxy Modelling with Next Generation Facilities}
\editors{M. Boquien, E. Lusso, C. Gruppioni, \& P. Tissera}
\begin{document}

\maketitle

\begin{abstract}
Fitting the multi-wavelength spectral energy distributions (SEDs) of galaxies is a widely used technique to extract information about the physical properties of galaxies. However, a major difficulty lies in the numerous uncertainties regarding almost all ingredients of the SED modeling of galaxies. The Bayesian methods provide a consistent conceptual basis for dealing with the problem of inference with many uncertainties. While the Bayesian parameter estimation method have become quite popular in the field of SED fitting of galaxies, the Bayesian model comparison method, which is based on the same Bayes' rule, is still not widely used in this field. With the application of Bayesian model comparison method in a series of papers, we show that the results obtained with Bayesian model comparison are understandable in the context of stellar/galaxy physics. These results indicate that Bayesian model comparison is a reliable and very powerful method for the SED fitting of galaxies.

\end{abstract}

\firstsection 
\section{Introduction}
As shown in a series of papers with the same title by Conroy and collaborators \citep{Conroy2009a,Conroy2010c,Conroy2010d}, there are many important uncertainties in the stellar population synthesis modeling.
To obtain the synthetic spectra of simple stellar populations (SSPs), we need to employ a library of stellar evolutionary tracks, a library of stellar spectra, and assumption about the stellar initial mass function (IMF), all of which suffer from some important uncertainties.
It is still challenging to model the complex effects of convection, rotation and a binary companion in stellar evolution \citep{PaxtonB2013a}.
Some not well understood phases of stellar evolution, such as the thermally pulsating AGB stars \citep{Marigo2007a,Marigo2008a,RosenfieldP2016a}, horizontal branch stars \citep{Heber2009a,CassisiS2013a}, and blue straggler stars \citep{BoffinH2014a}, have important contributions to the integrated light from a galaxy.
On the other hand, the empirical libraries \citep[e.g.,][]{Vazdekis2010a} of stellar spectra are limited in the coverage in the parameter space, possible errors in the estimation of physical parameters, finite SNR, while the theoretical ones \citep[e.g.,][]{Leitherer2010a} suffer from shortcomings of atomic and molecular data, inappropriate abundance ratios, non-LTE, etc.
Futhermore, the stellar initial mass functions (IMFs) may not be universal as traditionally assumed, but could be variable in the galaxy's formation history \citep{Kroupa2001a,BastianN2010a,Cappellari2012a}.
Consequently, the numerous uncertainties in the ingredients of the modeling have led to the diversity of stellar population synthesis (SPS) models\footnote{See \url{http://www.sedfitting.org/Models.html} for an incomplete list of these models.}.
Except for the SSPs given by SPS modelings, we need models about the star formation history (SFH) and chemical evolution history of the interstellar medium, which are even more uncertain.

\section{Bayesian model comparison}
Given the numerous uncertainties regarding almost all ingredients of the SED modelings of galaxies, we need to find a method to allow a reasonable comparison of the many possible assumptions about them.
While the Bayesian parameter estimation method has been widely used in the field of  SED fitting galaxies, the Bayesian model comparaison method, which is based on the same Bayes' rule as Bayesian parameter estimation, is still rarely used in this field.
A main reason for this situation is that the computation of Bayesian evidence, which is critical for Bayesian model comparison, is generally much more computationally expansive than the posterior sampling in Bayesian parameter estimation.
However, with the advent of state-of-the-art Bayesian inference tool, such as MultiNest \citep{Feroz2009a},  it has become possible to achieve efficient and robust computation of Bayesian evidence.
The Bayesian evidence can be thought as a quantified Occam's razor such that a model with a better balance between its goodness-of-fit to the data and its complexity will be more favored \citep{Gregory2005a}.

A major difficulty for the application of Bayesian model comparison method to the comparison of different SED modelings lies in the fact that the computation of the SED modeling itself could be even more computationally expansive. 
In \cite{Han2012a}, we have employed the artificial neural network (ANN) technique to allow a rather rapid computation of model SEDs.
With the building-up of a pre-released version of our BayeSED code, we have applied the Bayesian model comparison to the study of a sample of hyper-luminous infrared galaxies (HLIRGs) \citep{Ruiz2010a}.
We found that the more complicated Starburst+AGN model has larger Bayesian evidence than the simpler pure Starburst or pure AGN model.
This result supports the idea that HLIRGs represent an important phase in the formation and evolution of galaxies where both the star formation and SMBH accretion are very active.
In \cite{FanL2016a,FanL2016b}, we have applied the similar methods to the infrared SED decomposition of a sample of WISE-selected, hyperluminous hot dust-obscured galaxies.
We found that the more complicated AGN torus+Greybody model has larger Bayesian evidence than the simpler pure AGN torus or pure Greybody model for these galaxies.

In \cite{HanY2014a}, we have presented the first version of our BayeSED code.
The performance of BayeSED for the estimation of physical parameters, such as stellar mass and star formation rate, has been extensively tested with a mock sample of galaxies and the comparaison with the results obtained with the FAST code \citep{Kriek2009a}.
With BayeSED V1.0, we have presented the first Bayesian comparison of SPS models.
For a Ks-selected sample of 5467 low-z galaxies given by \cite{Muzzin2013a}, we found that the simpler BC03 model has larger Bayesian evidence than the M05 model.
However, this comparison of SPS models has some important limitations.
For example, we have only considered the most popular BC03 and M05 models, while there are many others.
The SFH of galaxies is fixed to the commonly used  Exp-dec form, and the dust attenuation law (DAL) is fixed to the Calzetti law.
Since the SFH and DAL of different galaxies could be very different, the above assumption about them could introduce serious bias to the results of Bayesian SPS model comparaison.

In \cite{HanY2019a}, we have presented a more comprehensive Bayesian Discrimination of SPS models, where 16 SSP models, 5 forms of SFH and 4 forms of DAL have been considered.
We have introduced a method to define the Bayesian evidence for the SED modeling of a sample of galaxies with a universal and fixed SSP, but object-dependent and free SFH and DAL.
Since the Bayesian evidence is defined for a sample of galaxies instead of an individual galaxy and the results are obtained with the marginalization over the different choices of SFH and DAL for different galaxies, we can achieve a much more robust comparison of SPS models.
We found that the most widely used BC03 has much larger Bayesian evidence than many other more updated SPS models, including the CB07 model from the same authors.
Our results indicate that the contribution of TP-AGB stars is still not properly considered in current SPS models.
On the other hand, we found that the version of BPASS V2.0 model \citep{EldridgeJ2008a} without binaries has a slightly larger Bayesian evidence than the BC03 models.
However, the version of BPASS V2.0 model with binaries actually has the lowest Bayesian evidence.
For the two models with the consideration of binaries, the Yunnan-II model \citep{Zhang2005a} is more favored than the BPASS V2.0 model.
Our results indicate that the effects of binaries are still not well considered in current SPS models, which is not surprising, given the uncertainties about the period and mass-ratio distributions of binary stars \citep{MoeM2017a} and other not well understood complex physics in binary stellar evolution.

\section{Conclusion}                              
With the MultiNest-like Bayesian inference tools and methods for rapid generation of model SEDs, it has become possible to reliably and efficiently compute the Bayesian evidence for a large set of SED models.
In a series of papers, we show that the results obtained with Bayesian model comparison are understandable in the context of stellar/galaxy physics.
So, we believe that Bayesian model comparison is a reliable and very powerful method, and it should be more extensively used in the field of SED fitting and the study of galaxies.
We have built the BayeSED code \footnote{\url{https://bitbucket.org/hanyk/bayesed/}} for this kind of studies.
The public version of the code is in a compiled binary format, which is generally faster than other interpreted language code (e.g. Python).
The code is directly runable on MAC/LINUX platform and does not require users to resolve the often annoying library dependency problems.
In the future, more and more sophisticated SED models will be added gradually to BayeSED.
The suggestions about adding new SED models and other new features, and the reports about bugs are very welcome.

\section{Acknowledgements}
This work is supported by the National Natural Science Foundation of China (NSFC, Grant Nos. 11773063, 11521303, 11733008), Chinese Academy of Sciences (CAS, no. KJZD-EW-M06-01) and Yunnan Province (Grant No. 2017FB007, 2017HC018).


\bibliography{IAU341_hanyk.bbl}

\begin{discussion}

\discuss{Joel Leja}{Given the many uncertainties in SED modeling, can you be sure that a model comparison is uniquely showing that that parameter is favored?}

\discuss{Yunkun Han}{No. When you want to answer questions about one specific parameter (e.g. SSP model) with Bayesian model comparison, other uncertainties (SFH, DAL) can be marginalized out to obtain robust answer. However, there are many possibilities for the other two. Indeed, to make the results of model comparison as robust as possible, we should make the selection of the other two as flexible as possible.} 
\discuss{David Rosario}{Can you distinguish between the model parameter and systematic issues with the dataset?}

\discuss{Yunkun Han}{Yes. Except for the normal physical parameters, we have  added the parameter $err_{\rm sys}^{\rm obs}$ to take into account the possible systematic issues within the dataset.} 

\end{discussion}

\end{document}